\documentclass{aa}
\usepackage[varg]{txfonts}
\usepackage{hyperref}
\hypersetup{colorlinks=true,linkcolor=blue,citecolor=blue,filecolor=blue,urlcolor=blue}

\usepackage{graphicx}
\graphicspath{{images/}}

\newcommand{\pc}{\mathrm{pc}}
\newcommand{\kpc}{\mathrm{kpc}}

\newcommand{\myr}{\mathrm{Myr}}
\newcommand{\msun}{M_\odot}

\newcommand{\mcl}{M_\mathrm{cl}}

\newcommand{\rh}{r_\mathrm{h}}

\newcommand{\trel}{t_\mathrm{rh}}
\newcommand{\tcc}{t_\mathrm{cc}}
\newcommand{\tcr}{t_\mathrm{cr}}
\newcommand{\tauv}{\tau_\mathrm{v}}
\newcommand{\fbin}{f_\mathrm{bin}}
\newcommand{\nbin}{n_\mathrm{bin}}

\defcitealias{pav_segr}{Paper~I}
\defcitealias{pav_telc}{Paper~II}

\begin{document}

\title{Primordial mass segregation of star clusters with primordial binaries}
\subtitle{}
\titlerunning{Primordial mass segregation of star clusters: The role of binary stars}

\author{V\'aclav Pavl\'ik \inst{\ref{asu},\ref{auuk},\ref{obs},}\thanks{\email{pavlik@asu.cas.cz}}}

\institute{Astronomical Institute of the Czech Academy of Sciences, Bo\v{c}n\'i~II~1401, 141~31~Prague~4, Czech Republic\label{asu}
\and Astronomical Institute of Charles University, V Hole\v{s}ovi\v{c}k\'ach 2, 180~00~Prague~8, Czech Republic\label{auuk}
\and Observatory and Planetarium of Prague, Kr\'alovsk\'a obora 233, 170~21~Prague~7, Czech Republic\label{obs}}

\authorrunning{Pavlík, V.}

\date{Received: January 13, 2020 / Accepted: April 29, 2020}

\abstract
{Observations of young star-forming regions suggest that star clusters are born completely mass segregated. These initial conditions are, however, gradually lost as the star cluster evolves dynamically. For star clusters with single stars only and a canonical initial mass function, it has been suggested that traces of these initial conditions vanish at a time $\tauv$ between 3 and $3.5\,\trel$ (initial half-mass relaxation times).}
{Since a significant fraction of stars are observed in binary systems and it is widely accepted that most stars are born in binary systems, we aim to investigate what role a primordial binary population (even up to 100\,\% binaries) plays in the loss of primordial mass segregation of young star clusters.}
{We used numerical $N$-body models similar in size to the \object{Orion Nebula Cluster} (\object{ONC}) -- a representative of young open clusters -- integrated over several relaxation times to draw conclusions on the evolution of its mass segregation. We also compared our models to the observed \object{ONC}.}
{We found that $\tauv$ depends on the binary star fraction and the distribution of initial binary parameters that include a semi-major axis, eccentricity, and mass ratio. For instance, in the models with 50\,\% binaries, we find $\tauv = (2.7 \pm 0.8)\,\trel$, while for 100\,\% binary fraction, we find a lower value $\tauv = (2.1 \pm 0.6)\,\trel$.
We also conclude that the initially completely mass segregated clusters, even with binaries, are more compatible with the present-day \object{ONC} than the non-segregated ones.}
{}

\keywords{methods: numerical, data analysis -- star clusters: individual (ONC) -- stars: formation, binaries}

\maketitle

\section{Introduction}

As a star cluster dynamically evolves, mass segregation establishes, and we observe a~general tendency of clusters to evolve towards an even higher degree of mass segregation \cite[e.g.][]{chandrasekhar,chandra_vonNeumann1,chandra_vonNeumann2}. Recent observations of the Serpens South star-forming region performed with \textsl{ALMA} by \citet{plunkett}, however, suggest that young clusters are already born completely mass segregated. In \citet[][hereafter Paper I]{pav_segr}, we investigated this primordial mass segregation and were the first to point out that it is unexpectedly gradually lost because of two-body encounters that lead to energy equipartition. We then concluded that the degree of mass segregation of a non-segregated and a completely segregated system should gradually settle at a similar level. Hence, both initial conditions become observationally indistinguishable (i.e.\ the primordial difference vanished) after some time designated as $\tauv$. In $N$-body models with single stars only, we estimated this time to be $3 < \tauv / \trel < 3.5$ \citep[where $\trel$ stands for the half-mass two-body relaxation time; cf.][]{spitzer_hart_relax}.

It is argued, for example, by \citet{kroupa95a} and \citet{go_kr05} that most (perhaps all) stars are preferentially born in pairs. Observations report 57\,\% of G-dwarfs \citep{duq_may91,rag_etal10}, 45\,\% of K-dwarfs \citep{may_etal92}, and 42\,\% of field, that is old, M-dwarfs \citep{fi_ma92} are present in binaries. And the binary fraction increases with the stellar mass. This dependency on primary mass results if all stars form as binaries, which are dynamically processed in embedded clusters of various masses \citep[][and their Fig.~6]{marks_kroupa11,thies_etal15}. Hence, we extended the work presented in \citetalias{pav_segr} by studying the evolution of mass segregation in star clusters that include primordial binaries in \cite[][hereafter Paper II]{pav_telc}, using several realisations of a model with 50\,\% binaries. In this work, we go even further with higher number of realisations for better statistical description, different binary star distributions, and also binary fractions up to 100\,\%.

\section{Models}

\begin{figure*}
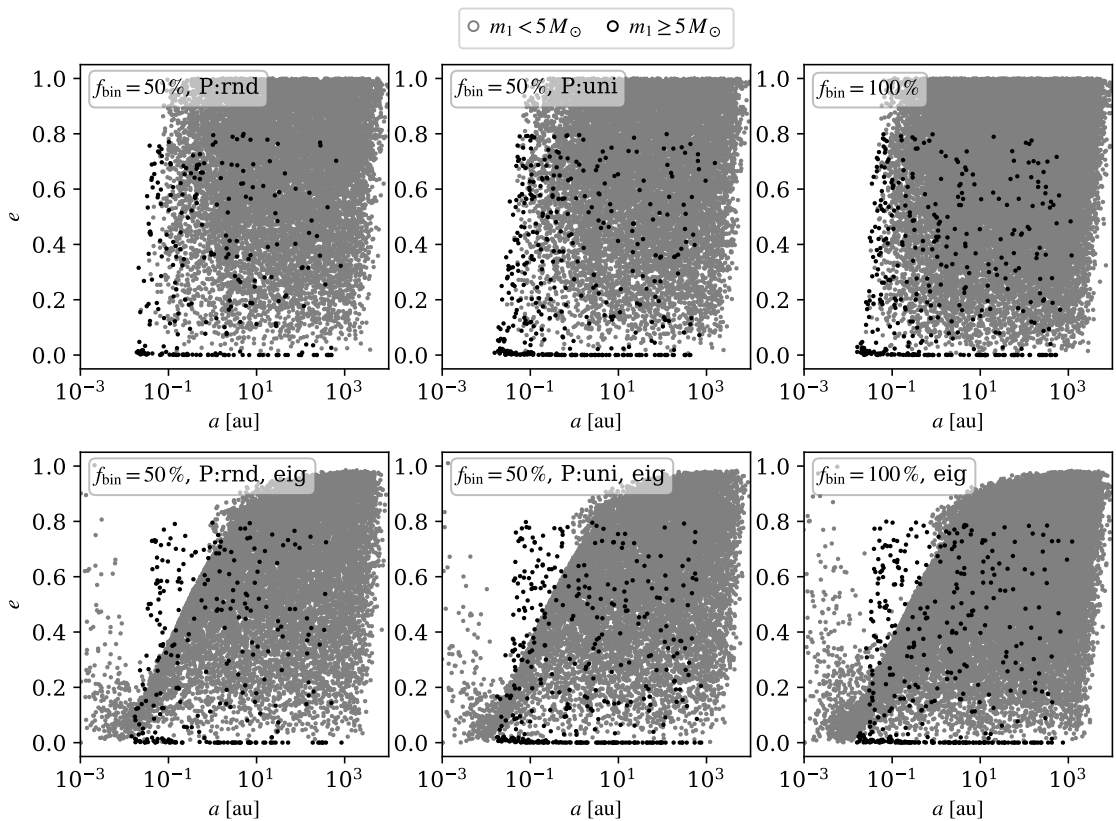

        \centering
        \includegraphics[width=.82\linewidth]{{{binseg_par}}}

        \includegraphics[width=.82\linewidth,trim={0 0 0 1cm},clip]{{{binseg_par-eig}}}
        \caption{Initial semi-major axis and eccentricity of the binary populations used in this work, distinguished by the primary component mass -- either $m_1 < 5\,\msun$ (grey) or $m_1 \geq 5\,\msun$ (black). Twenty realisations from models with and without eigenevolution are used, while segregated and non-segregated models are combined in these plots.}
        \label{fig:ic}
\end{figure*}

\begin{table}[!htb]
  \centering 
  \caption{Initial parameters of the star cluster models.}
  \begin{tabular}{lccc}
    \hline
    $\fbin$          & 50\,\% (\texttt{P:rnd})   & 50\,\% (\texttt{P:uni})   & 100\,\% \\
    \hline
    $N$              & 2404    & 2404   & 2404   \\
    $\mcl\ [\msun]$  & 1318.1  & 1318.1 & 1318.1 \\
    $\rh\ [\pc]$     & 0.258   & 0.263  & 0.262  \\
    $\tcr\ [\myr]$   & 0.219   & 0.221  & 0.221  \\
    $\trel\ [\myr]$  & 3.67    & 3.77   & 2.68   \\
    \hline
    \texttt{seg} &&& \\
    $m_{\rm 5}\ [\msun]$                              & 100.5  & 116.8 & 113.7  \\
    $\rho_{\rm c}\ [\times 10^3 \frac{\msun}{\pc^3}]$ & 31.3   & 31.7  & 29.6   \\
    $\rho_{\rm h}\ [\times 10^3 \frac{\msun}{\pc^3}]$ & 9.7    & 9.7   & 9.6   \\
    \hline
    \texttt{non} &&& \\
    $m_{\rm 5}\ [\msun]$                              & 2.61   & 3.84  & 2.74   \\
    $\rho_{\rm c}\ [\times 10^3 \frac{\msun}{\pc^3}]$ & 33.3   & 33.9  & 31.3   \\
    $\rho_{\rm h}\ [\times 10^3 \frac{\msun}{\pc^3}]$ & 10.3   & 10.1  & 9.7   \\
    \hline
    \# runs                                           & 85     & 50    & 80     \\
    \# runs (\texttt{eig})                            & 160    & 80    & 90     \\
    \hline
  \end{tabular}
  \label{tab:models}
  \tablefoot{The columns represent initial binary fractions and pairings. The top part shows: the number of stars, mass of the cluster, half-mass radius, crossing time, and median relaxation time of the models. The middle two parts show: the mass of the five innermost stars, the density enclosed in the central 0.1\,pc and in the half-mass radius for both \texttt{seg} and \texttt{non} initial conditions. The mean values for each model are given. The bottom part shows the number of runs, which is the same for \texttt{seg} and \texttt{non} models, but differs for models with and without eigenevolution.}
\end{table}

We prepared an ensemble of isolated $N$-body models with 2.4k stars in the mass range from 0.08 to $50\,\msun$ \citep[comparable to the sources in the \object{Orion Nebula Cluster}, \object{ONC},][]{onc_data} with the optimally sampled canonical IMF \citep{kroupa,kroupa_etal13}, where the highest stellar mass is determined from the maximum stellar-mass-to-cluster-mass relation \citep{max_stellar_mass,pflamm_old_stars}\footnote{The mass range of the initial conditions is the same as in \citetalias{pav_segr}.}\!.
All clusters started as Plummer models \citep{plummer,andre_etal}, where the half-mass radius was determined from the birth radius to embedded cluster mass relation \citep{marks_kroupa}. We evolved them for several relaxation times using the state-of-the-art numerical integrator \texttt{nbody6} \citep{aarseth}\footnote{Public version from May 30, 2016.}.

\begin{figure*}
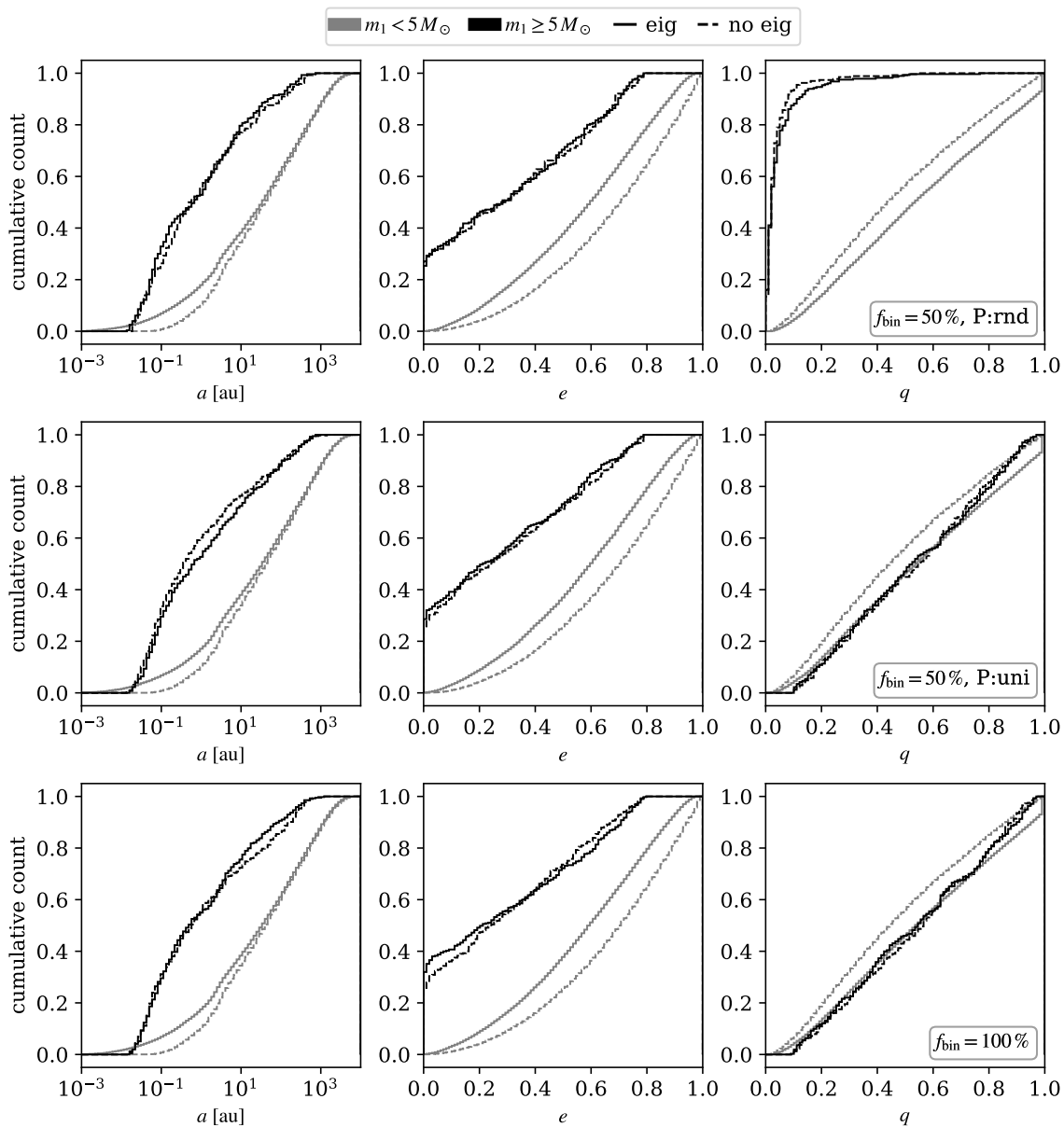

        \centering
        \includegraphics[width=.82\linewidth]{{{bin_seg_0-binseg_hist_aeq}}}

  \includegraphics[width=.82\linewidth,trim={0 0 0 1cm},clip]{{{bin_seg_3-binseg_hist_aeq}}}

        \includegraphics[width=.82\linewidth,trim={0 0 0 1cm},clip]{{{bin_100-binseg_hist_aeq}}}
        \caption{Initial distributions of the semi-major axis, eccentricity and mass ratio of the binary populations used in this work. Two mass groups are distinguished according to the primary component mass: $m_1 < 5\,\msun$ (grey) or $m_1 \geq 5\,\msun$ (black). Models with (solid line) and without (dashed line) eigenevolution are compared. Models with and without primordial mass segregation (ten realisations each) are combined in each curve.}
        \label{fig:ic_hist}
\end{figure*}

Similarly to \citetalias{pav_segr}, we set up two extreme degrees of primordial mass segregation according to a~method of \citet{baumgardt_segr} -- none or complete, for instance, the mass segregation parameter is $S=0$ or $S=1$, respectively. The primordially mass-segregated models and their parameters are labelled \texttt{seg}, the non-segregated ones are labelled \texttt{non}. The models' physical properties are summarised in Tab.~\ref{tab:models}.

In this work, two initial binary populations were analysed: (i) a conservative 50\,\% (i.e.\ 601 binary stars in total) and (ii) 100\,\% stars in binaries. In the models with 50\,\% binaries, two pairing methods were employed \citep[cf.][]{mcluster}\footnote{Public version from Jul 6, 2018. \url{https://github.com/ahwkuepper/mcluster}}: (a) random pairing across the whole mass range (labelled \texttt{P:rnd} in the following plots), and (b) pairing based on a uniform mass ratio distribution ($0.1<q<1.0$) for all stars above $5\,\msun$ and random pairing of the remaining stars up to the desired percentage (labelled \texttt{P:uni}, as well as in \citetalias{pav_telc}). In the case of the models with 100\,\% binaries, only the uniform pairing method (described above) was used.
The semi-major axes were distributed according to \citet{sana_etal12} and \citet{oh_etal15} distributions for stars with $m > 5\,\msun$ (with the periods $P$ in days such that $\log_{10}{P} \geq 0.15$) and according to \citet{kroupa95a} for lower mass stars (with $\log_{10}{P} \geq 1$).
Eccentricity distribution of high-mass systems was taken from \citet{sana_evans11} and was thermal for low-mass stars \citep[cf.][]{heggie,duq_may91,kroupa08}. The low-mass short-period binaries were initialised in two different ways: (i) in the models labelled \texttt{eig}, we adopted \cite{kroupa95} pre-main sequence eigenevolution, and (ii) in the models without such a label, we did not. Results of both of these initial conditions are also compared with respect to the cluster's mass segregation. For reference, the initial binary population parameters (semi-major axis, eccentricity, and mass ratio) are plotted in Figs.~\ref{fig:ic}~\&~\ref{fig:ic_hist} for all models.

Each model in Tab.~\ref{tab:models} is simulated many times (see the bottom two rows) for each combination of the initial parameters using different random seeds to acquire good statistics.
The models presented are still idealised in the sense that the stellar evolution is suppressed, the interstellar gas (and therefore gas expulsion) is not considered, and the star clusters are isolated. We discuss the effects of these features in Appendix~\ref{ap:discussion}.

The initial comparison of different pairing methods with the same fraction of binaries (i.e.\ models \texttt{P:rnd} and \texttt{P:uni}) in the high-mass range (primary component has $m_1 > 5\,\msun$; i.e.\ black lines in Fig.~\ref{fig:ic_hist}) reveals that the \texttt{P:rnd} model has no equal-mass binary stars and instead produces $\approx 80\,\%$ of systems ($\approx 70\,\%$ after eigenevolution) where the primary component is at least 20 times more massive than the secondary ($q < 0.05$). Generally, when primordial binaries are included in the models, they are able to speed up mass segregation in the non-segregated clusters because they are viewed by the rest of the stars as more massive bodies. In this particular case, where the random pairing of high-mass stars with very low-mass stars predominantly occurs, the dynamical impact of binaries with high-mass primary components is not going to be much higher than if these components were just single stars.

Concerning the models with and without eigenevolution, there is almost no difference in the initial distribution of the high-mass binaries. On the other hand, the lower-mass binaries (where the primary component has $m_1 < 5\,\msun$) do change their initial properties after the pre-main sequence eigenevolution. (i) There is a higher number of close binary systems (compare the plots in the top and bottom rows of Fig.~\ref{fig:ic}, and the plotted dashed and solid grey lines in the left column of Fig.~\ref{fig:ic_hist}). (ii) The eccentricities become more uniformly distributed between 0 and 1 (see the middle column of plots in Fig.~\ref{fig:ic_hist}). (iii) The mass ratios shift towards unity, thus, more equal-mass binaries are created (see the spike at $q = 1$ in the right column of plots in Fig.~\ref{fig:ic_hist}). These binaries could potentially speed up the mass segregation in the non-segregated models.

\section{Results}

\subsection{Mass segregation}

\begin{figure*}
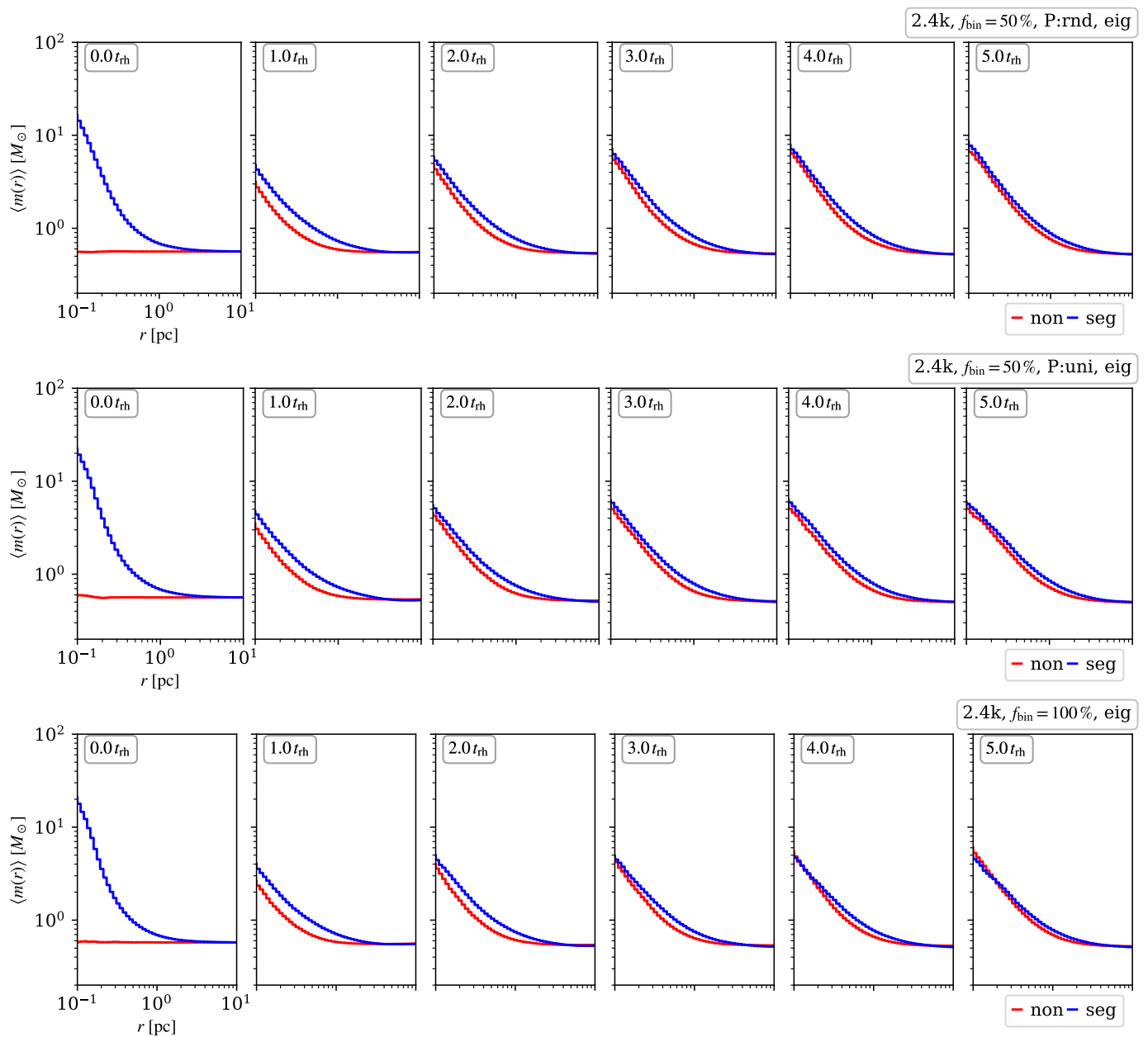

        \centering
        \includegraphics[width=\linewidth]{{{bin_seg_0-x-mean_mass_bin_10.0_p0_0-29-2}}}

        \includegraphics[width=\linewidth]{{{bin_seg_3-x-mean_mass_bin_10.0_p0_0-29-2}}}

        \includegraphics[width=\linewidth]{{{bin_100-x-mean_mass_bin_10.0_p0_0-21-2}}}
        \caption{Evolution of radial distribution of the mean stellar mass in the cluster models with eigenevolution (red lines are for non-segregated models and blue lines for segregated models). Each row of plots corresponds to one model (top and middle have 50\,\% initial binaries but different pairing methods; the bottom row is with 100\,\% initial binaries). Time increases from left to right in the units of the initial half-mass relaxation time.}
        \label{fig:mean_mass}
\end{figure*}

\begin{figure*}
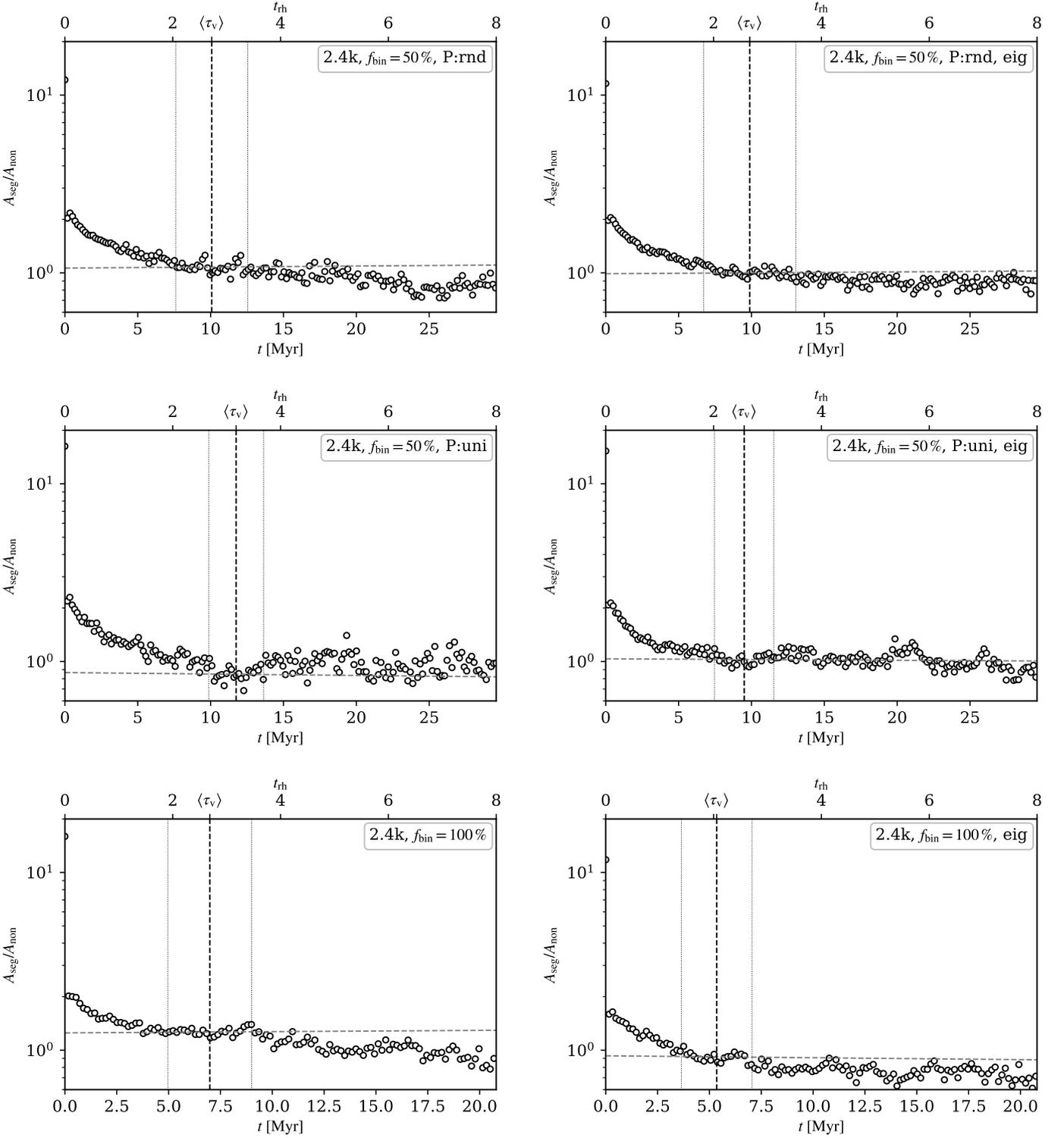

        \centering
        \includegraphics[width=.49\linewidth]{{{bin_seg_0-solo-seg_par_bin_0.1-10.0-2}}}\hfill
        \includegraphics[width=.49\linewidth]{{{bin_seg_0-solo-seg_par_bin_0.1-10.0-eig-2}}}

        \includegraphics[width=.49\linewidth]{{{bin_seg_3-solo-seg_par_bin_0.1-10.0-2}}}\hfill
        \includegraphics[width=.49\linewidth]{{{bin_seg_3-solo-seg_par_bin_0.1-10.0-eig-2}}}

        \includegraphics[width=.49\linewidth]{{{bin_100-solo-seg_par_bin_0.1-10.0-2}}}\hfill
        \includegraphics[width=.49\linewidth]{{{bin_100-solo-seg_par_bin_0.1-10.0-eig-2}}}
        \caption{Evolution of the ratio given by Eq.~\eqref{eq:area} in time of the models used. The plots on the right-hand side are with short-period binary eigenevolution. The black vertical dashed line and the value $\langle\tauv\rangle$ represent the mean time when the slope of the data points became flat, the thinner vertical lines are the calculated $1\sigma$ uncertainties. The horizontal fit in the corresponding time window is indicated by a grey dashed line.}
        \label{fig:tauv}
\end{figure*}

Our models with binaries (both 50\,\% and 100\,\%) evolve in a similar way to the single star models presented in \citetalias{pav_segr}. The difference in the \texttt{seg} and \texttt{non} clusters is very high initially. The primordially fully mass-segregated clusters, however, start to lose their initial ordering gradually due to random two-body encounters between massive stars, especially in the core region, which can efficiently eject stars out of the cluster \citep[cf.][]{oh_etal15,oh_kroupa16,kroupa_onc,long_etal18}. At the same time, the clusters without initial mass segregation do quickly establish it dynamically. After a period of time, both models settle almost at the same level of mass segregation. As in \citetalias{pav_segr}, we investigated the evolution of mass segregation using the spatial distribution of mass. We read the mean mass comprised in concentric spheres around the centre of the cluster \citep[the density centre according to a method of][implemented in \texttt{nbody6}, is taken]{casertano_hut}, as plotted in Fig.~\ref{fig:mean_mass} for each model, and calculated an integral parameter
\begin{equation}
        \label{eq:area}
        A = \sum_{k = 1}^{\nbin}{\frac{\langle m(r_k) \rangle}{\Delta r_k}} \,,
\end{equation}
with the mean mass in the $k$-th bin at radius $r_k$
\begin{equation}
        \langle m(r_k) \rangle = \frac{\sum_{i = 1}^k{m_i}}{\sum_{i = 1}^k{n_i}} \,,
        \vspace{-5pt}
\end{equation}
where $n_i$ and $m_i$ are the number of stars, and their total mass in an $i$-th bin, respectively. $\Delta r_k$ is the width of the $k$-th bin, thus the thickness of the spherical shell that is being added, and $\nbin$ is the total number of bins. The bins here are logarithmically equidistant, and, in particular, we chose $r_1 = 0.1\,\pc$, $r_{\nbin} = 10\,\pc$ and $\nbin = 50$; the lower boundary is approximately the core radius, while the upper boundary initially contains almost the whole cluster (see Fig.~\ref{fig:lagr}) and would be equivalent to the tidal radius (see Appendix~\ref{ap:discussion} and Fig.~\ref{fig:lagr_tide}).

The time when the difference of the initial conditions stabilises is estimated in each model from the ratio $A_\texttt{seg}/A_\texttt{non}$. We fitted this ratio by a linear function over a moving window of various lengths from $0.5$ to $1.5\,\trel$ and looked for the moment when the slope first became horizontal (due to the noisiness of the data, we allowed for the slope to incline up to $\pm 10^{-4}$ from zero). The median value of these times from all the used windows is then used as $\tauv$.
Without any constraints on the vertical position of the fit, the plateau around $\tauv$ appears always near $A_\texttt{seg}/A_\texttt{non} \approx 1$, which signifies that the primordial mass segregation truly vanished around that time, and these clusters would be observationally indistinguishable from each other.
Since in all models (with or without eigenevolution) the slopes after $\langle\tauv\rangle$ are near flat or have only a shallow slope (for example, in comparison to the first relaxation time) the clusters seem to evolve synchronously. This may also be verified with Fig.~\ref{fig:mean_mass}, where the evolution of each cluster radial profile is plotted (only models with eigenevolution are shown, because the models without it are very similar, which can also be verified from the evolution plotted in Fig.~\ref{fig:tauv}).

Systems with 50\,\% binaries and random pairing (\texttt{P:rnd}) yield $\langle \tauv \rangle \approx 2.7\,\trel$ with or without the pre-main sequence eigenevolution of the short-period binaries (see the top panels of Fig.~\ref{fig:tauv}). The core of the primordially mass-segregated cluster is populated by binaries composed of stars with more than $5\,\msun$ paired with 10 to 20 times less massive companions (for the mass ratios, see the top right plot in Fig.~\ref{fig:ic_hist}; for the initial core population, compare the blue curves at $t=0\,\trel$ in Fig.~\ref{fig:mean_mass}). These binaries, which are effectively `single' stars, are not as efficient in ejecting themselves out of the core as the equal-mass high-mass binaries would be, so they reside there longer. From the point of view of the primordially non-segregated model, these `single' stars are scattered all over the cluster, but also mainly around the core region. As they are still among the most massive bodies, they tend to segregate first, and because of their inefficiency to eject themselves out of the cluster afterwards, they partially halt further mass segregation of lower mass binaries. For the exact time needed to eject the most massive stars and how it compares to the other models, see the curves in Fig.~\ref{fig:top_mass}; and also compare the curves in Fig.~\ref{fig:mean_mass}, for example, at $t=5\,\trel$ (it stays higher in the \texttt{P:rnd} model than in the \texttt{P:uni}, meaning that these massive stars are still present in the core). Therefore, even though the pre-main sequence eigenevolution produces an excess of equal-mass low-mass binaries that are able to speed up the mass segregation of the non-segregated clusters (as they do in the following \texttt{P:uni} model), their effect is suppressed in the \texttt{P:rnd} model. Consequently, we cannot see any difference in the evolution of mass segregation between the \texttt{P:rnd} models with and without eigenevolution.

The systems with 50\,\% binaries and \texttt{P:uni} pairing seem to evolve from their initial mass segregation faster when the eigenevolution is accounted for. This could be due to the early ejection of high-mass binaries (see, e.g. Fig.~\ref{fig:top_mass}) and a higher number of low-mass equal-mass binaries that are present in the eigenevolved models. Such systems act in the same way as twice as massive single stars and, hence, they help the non-segregated clusters segregate faster. Yet, within their $1\sigma$ uncertainties, the resulting times $\tauv$ still seem to be comparable (see the middle panels of Fig.~\ref{fig:tauv}). We note that the value of $\tauv$ obtained in \citetalias{pav_telc} for \texttt{P:uni} model with only a handful of realisations holds even here, where more realisations are analysed.

When comparing both \texttt{P:rnd} and \texttt{P:uni} models with or without eigenevolution, Fig.~\ref{fig:tauv} shows a much milder and longer decrease of the ratio $A_\texttt{seg}/A_\texttt{non}$ in the \texttt{P:rnd} model, although its fitted value of $\tauv$ happens to be smaller (at least in the model without eigenevolution). The steeper decrease towards $A_\texttt{seg}/A_\texttt{non} \approx 1$ in the \texttt{P:uni} model may be caused again by a faster dynamical evolution of the non-segregated clusters due to an early dynamical influence of the equal-mass low-mass binaries.

In the case of the clusters with 100\,\% binaries, the time of vanishing of the initial mass segregation is consistent within the uncertainties with the models containing just 50\,\% binaries. In particular $\langle \tauv \rangle \approx 2.1\,\trel$ (with eigenevolution) and $\langle \tauv \rangle \approx 2.7\,\trel$ (without eigenevolution), see the bottom panels of Fig.~\ref{fig:tauv} and Tab.~\ref{tab:tauv}.

\begin{table}
  \centering 
  \caption{Resulting time $\tauv / \trel$.}
  \begin{tabular}{ccc}
    \hline
    eigenevolution           & without       & with (\texttt{eig}) \\
    $\fbin$ &&\\
    \hline
    50\,\% (\texttt{P:rnd})  & $2.7 \pm 0.7$ & $2.7 \pm 0.8$ \\
    50\,\% (\texttt{P:uni})  & $3.2 \pm 0.5$ & $2.6 \pm 0.6$ \\
    100\,\%                  & $2.7 \pm 0.8$ & $2.1 \pm 0.6$ \\
    \hline
  \end{tabular}
  \label{tab:tauv}
  \tablefoot{The $1\sigma$ uncertainties are given.}
\end{table}

The values of $\tauv$, which are summarised in Tab.~\ref{tab:tauv}, are generally smaller in comparison to the single-star models in \citetalias{pav_segr} (i.e.\ $3 < \tauv / \trel < 3.5$), but not significantly so, with the exception of the \texttt{eig} model containing 100\,\% binaries. Two factors play a role here. While in the clusters without primordial binary population binaries start to form as late as during the core collapse \citep[cf.][]{fujii_pz,pavl_subr}, here we already injected them in our models at the beginning of the integration. Binary stars then lead to a faster dynamical evolution, as they act as more massive bodies. The opposing factor is that most of the binaries that are present are initially soft (especially in the models without eigenevolution) so they very quickly disrupt, leaving their components as single stars behind. Therefore, the time $\tauv$ is expected to be smaller but not significantly so, for example, approximately by the time the single star models from \citetalias{pav_segr} needed to reach the core collapse (i.e.\ $0.5\,\trel$), which is what our results indicate. We also note that the core collapse of the models presented here happens only a fraction of the relaxation time sooner (i.e.\ $\tcc \approx 0.25\,\trel$, estimated from the minimum of the inner Lagrangian radius, see Fig.~\ref{fig:lagr}).

To summarise, if we compare different pairing schemes for the same percentage of binaries (i.e.\ \texttt{P:rnd} and \texttt{P:uni} models), the deduced values of $\tauv$ are almost the same. The biggest difference is visible in the models with 50\,\% \texttt{P:uni} or 100\,\% binaries where the value of $\tauv$ is on average about 20\,\% smaller if the pre-main sequence eigenevolution is accounted for. Higher numbers of close binary systems and more equal-mass binaries are generated initially in the eigenevolved models (compare the top and bottom panels in Fig.~\ref{fig:ic}, and the left and right column of plots in Fig.~\ref{fig:ic_hist}). Such binaries are harder to disrupt and could potentially speed up the mass segregation of the initially non-segregated clusters, thus shortening the time $\tauv$. However, based on the uncertainties of the fitted value $\tauv$, this difference may not have any statistical importance. Studying models with an even higher fraction of close binaries with the mass ratio biased towards unity could resolve this.

\subsection{The Orion Nebula Cluster}

\begin{figure*}
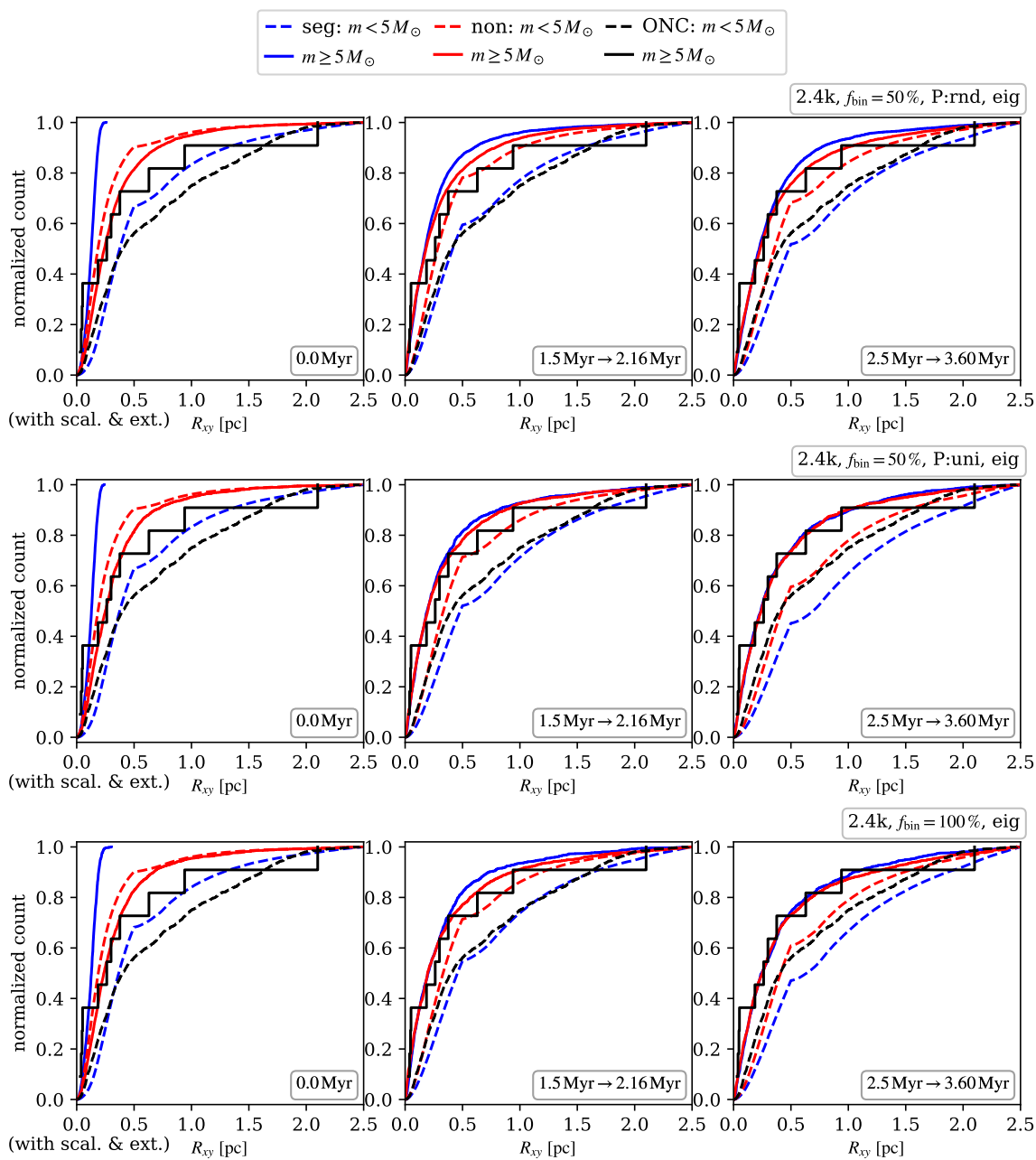

        \centering
        \includegraphics[width=.42\linewidth]{{{label_cum_rad}}}

%       \vspace*{10pt}
        \includegraphics[width=.82\linewidth]{{{bin_seg_0-binseg_cum_rad_merged_avg_extscl_mass_2.5-eig}}}

%       \vspace*{10pt}
        \includegraphics[width=.82\linewidth]{{{bin_seg_3-binseg_cum_rad_merged_avg_extscl_mass_2.5-eig}}}

%       \vspace*{10pt}
        \includegraphics[width=.82\linewidth]{{{bin_100-binseg_cum_rad_merged_avg_extscl_mass_2.5-eig}}}
        \caption{Comparison of average radial profiles of the projected models of segregated (blue) and non-segregated clusters (red) with the real \object{ONC} (black). Only the models with pre-main sequence eigenevolution are plotted. Three time frames are shown, and two mass groups are compared in each model ($m\geq 5\,\msun$ with a solid line and $m<5\,\msun$ with a dashed line). The $p$-values for these groups are plotted in Fig.~\ref{fig:onc}. All three models are modified by artificial extinction and scaling of the low-mass stars along the $y$-axis, as discussed in \citetalias{pav_segr}.}
        \label{fig:onc_radial_extscl}
\end{figure*}

\begin{figure}
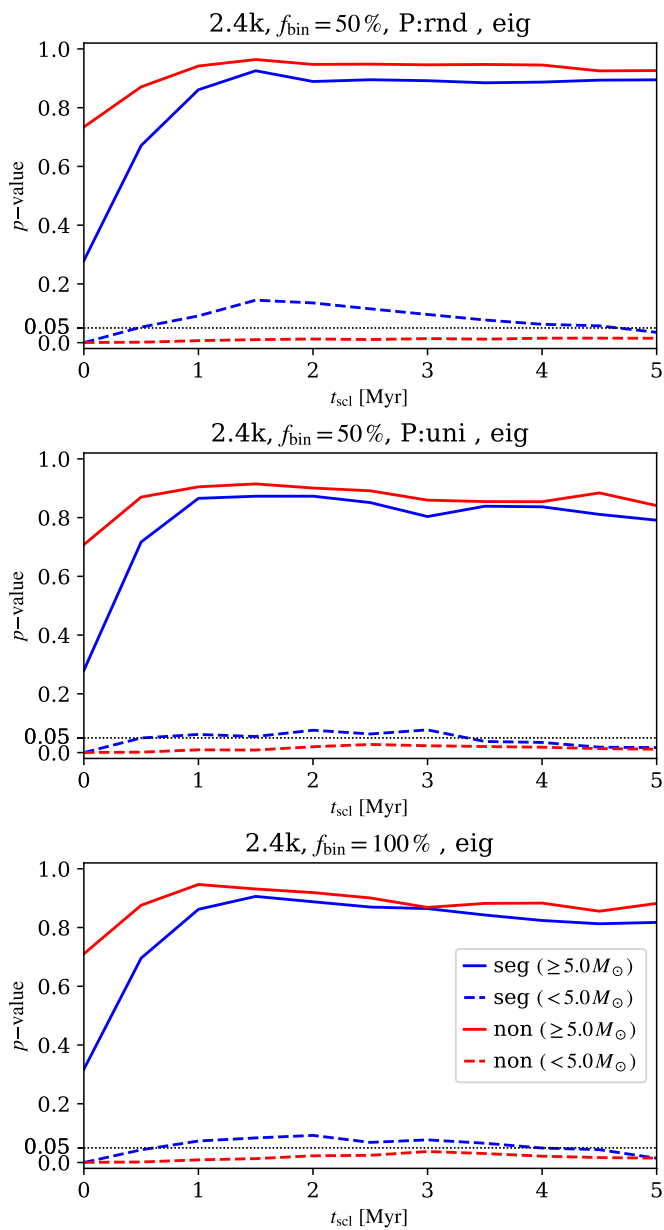

        \centering
        \includegraphics[width=\linewidth]{{{bin_seg_0-onc_ks_extscl_m5_2.5-eig}}}
        \includegraphics[width=\linewidth]{{{bin_seg_3-onc_ks_extscl_m5_2.5-eig}}}
        \includegraphics[width=\linewidth]{{{bin_100-onc_ks_extscl_m5_2.5-eig}}}
        \caption{Results of KS test between the \object{ONC} data and models with 50\,\% binaries (top and middle panels) and with 100\,\% binaries (bottom panel). Only models with scaling and extinction are shown, thus, the time is also in scaled units.}
        \label{fig:onc}
\end{figure}

As in \citetalias{pav_segr} and \citetalias{pav_telc}, we also tested which initial conditions are more compatible with the observed \object{ONC}. The present day \object{ONC} possesses about 8.5\,\% binaries \citep{reipurth}, but according to the simulations of \citet{king_onc}, it could even have had 73\,\% binaries initially.

From now on, we only use the models with pre-main sequence eigenevolution, as they give a more realistic population of binary stars (e.g.\ in the models without eigenevolution, the low-mass binaries with small semi-major axes are missing, and an artificial truncation is present). The radial distribution of stars in two mass bins (divided by $5\,\msun$) was calculated at different times and then compared to the same mass bins derived from the observational data in \citet{onc_data}. At each time, we provided the $p$-value of the Kolmogorov--Smirnov (KS) test. The distribution of stars of mass $m>5\,\msun$ is the same in the models as in the observed \object{ONC}. Low-mass stars, however, have significant differences -- taking the model's raw data showed no similarity with the data. This is mainly due to interstellar extinction \citep{scandariato_et_al} and geometry of the cluster \citep[elongation beyond 0.5\,pc from the core along the gaseous filament in which the \object{ONC} resides; cf.][]{hillenbrand_hartmann}. By modifying the data to account for both of these features (as in \citetalias{pav_segr}, Sect.~4), the primordially mass-segregated models reached the KS test $p>0.05$ between 1 and 3\,Myr, which is equivalent to the current age of the \object{ONC} \citep[2.5\,Myr;][]{hillenbrand,palla_stahler}. In the case of the initially non-segregated models, none are compatible with the present-day \object{ONC}, not even with scaling and extinction included, although the model with 100\,\% binaries was closely approaching the desired $p$ value at 3\,Myr. At three time frames, we plot the average radial distributions of stars in Fig.~\ref{fig:onc_radial_extscl} for visual comparison. The time evolution of the $p$-value of these modified models is plotted in Fig.~\ref{fig:onc}, the unmodified models (\texttt{seg} and \texttt{non}), which yield zero or almost zero everywhere in the low-mass range, are not included.
These results suggest that the \object{ONC} was likely completely mass-segregated at birth.

\section{Conclusions}

In this paper, we extend the works of \citet[][Paper~I]{pav_segr} and \citet[][Paper~II]{pav_telc}. We investigated the role of the primordial binary star population on mass segregation in star clusters of the size of the \object{ONC} using models with 50\,\% and 100\,\% initial binary stars.

In the models with 50\,\% initial binaries, we see no difference in the initial pairing when studying the evolution of mass segregation. We may also conclude that even with binaries, the models evolve in a similar fashion to the models with only single stars, yet they do so somewhat faster, which could be due to a faster dynamical evolution caused by the presence of hard binaries straight from the beginning. The loss of primordial mass segregation was milder in the models where pairing was random for all stars (\texttt{P:rnd}) in comparison to the models with a uniform distribution of mass ratio for binaries with stars above $5\,\msun$, and random only in the low-mass range (\texttt{P:uni}). Yet when the models were initialised with a pre-main sequence eigenevolution of short-period binaries, this apparent difference in the initial evolution partially faded. The time when the primordially mass-segregated and non-segregated models became observationally indistinguishable (i.e.\ the difference between different initial conditions vanishes) is $\langle\tauv\rangle = 2.7 \,\trel$ (\texttt{P:rnd,eig}) or $\langle\tauv\rangle = 2.6 \,\trel$ (\texttt{P:uni,eig}) for the models with pre-main sequence binary eigenevolution, and $\langle\tauv\rangle = 2.7 \,\trel$ (\texttt{P:rnd}) or $\langle\tauv\rangle = 3.2 \,\trel$ (\texttt{P:uni}) for those without it. However, we are still within the $1\sigma$ uncertainties of the single-star model from \citetalias{pav_segr} where the value was $\langle\tauv\rangle = 3.3\,\trel$.

In the models with 100\,\% binaries, we can also see that the initial difference of mass segregation disappeared. Again, the time needed was shorter than in the single-star clusters ($\langle\tauv\rangle = 2.1\,\trel$ with eigenevolution and $\langle\tauv\rangle = 2.7\,\trel$ without it).

All models with eigenevolution, which undoubtedly better represent the cluster binary population, also ended up better erasing the differences of primordial mass segregation than models without it. The fitted values of $\tauv$ from all models are mutually within their $1\sigma$ uncertainties, thus, there may be no statistical significance between the evolution of mass segregation in the presented models. However, on average, we can see that the models with pre-main sequence eigenevolution tend to erase the initial differences in mass segregation faster, for instance,\ the time $\tauv$ is approximately 20\,\% smaller than in the models without eigenevolution. Such a behaviour would be expected, because the pre-main sequence eigenevolution produces more equal-mass low-mass binaries that can speed up the mass segregation of the initially non-segregated clusters once the heavier binaries eject themselves from the cluster core. We note that models containing larger numbers of stars, and higher abundances of hard and equal-mass binaries (both high-mass and low-mass) are needed to fully determine this. Since the distribution of the initial binary parameters could have a stronger influence on the final results, it would also be important to follow the constrains from observations of semi-major axes, mass ratios, or metallicity \citep[see e.g.][]{2017ApJS..230...15M,2017MNRAS.471.2812B,2020arXiv200409525D}.

We also compared our models to the present-day Orion Nebula Cluster. When we also account for the interstellar extinction and physical elongation of the cluster, our results suggest that the \object{ONC} was probably primordially mass segregated. Models without primordial mass segregation are only compatible with the \object{ONC} for the high-mass stars, but never in the low-mass range. Further modelling including stellar evolution, elongation along the molecular cloud, and gas expulsion are, however, still needed to support this.

\begin{acknowledgements}
This study was supported by Charles University (grant SVV-260441), by the Czech Academy of Sciences (project RVO:67985815) and by the Czech Science Foundation (project of Excellence No.~18-20083S).
Computational resources were provided by the CESNET LM2015042 and the CERIT Scientific Cloud LM2015085, under the programme ``Projects of Large Research, Development, and Innovations Infrastructures''.
The author also appreciates discussion with the participants of the ``($M$+4)$^\mathrm{th}$ Aarseth $N$-body Meeting'' and especially the comments from Pavel Kroupa. The author feels indebted to an anonymous referee for their comments and suggestions.
\end{acknowledgements}

\bibliographystyle{aa}
\bibliography{37490-20}

% \clearpage
\appendix
% \onecolumn
\section{Discussion of the initial conditions}
\label{ap:discussion}

In this work, we simulated idealised star clusters, which means that the clusters are isolated, without intracluster gas, and stars were considered point masses without stellar evolution, although an IMF was used. Here, we further discuss how  these initial conditions might affect the results presented.

Using a point-mass approximation in star clusters (i.e.\ within the framework of the $N$-body problem) is reasonable since the stellar radii are much smaller than the separation of stars -- tidal effects are therefore negligible. Very close random encounters, which could lead to stellar collisions did not occur in the simulations either. The only issue is that a few binaries were initialised as very tight ones (i.e.\ possible candidates for stellar collisions), but their components would not be able to merge as they are treated as point masses.

\begin{figure}
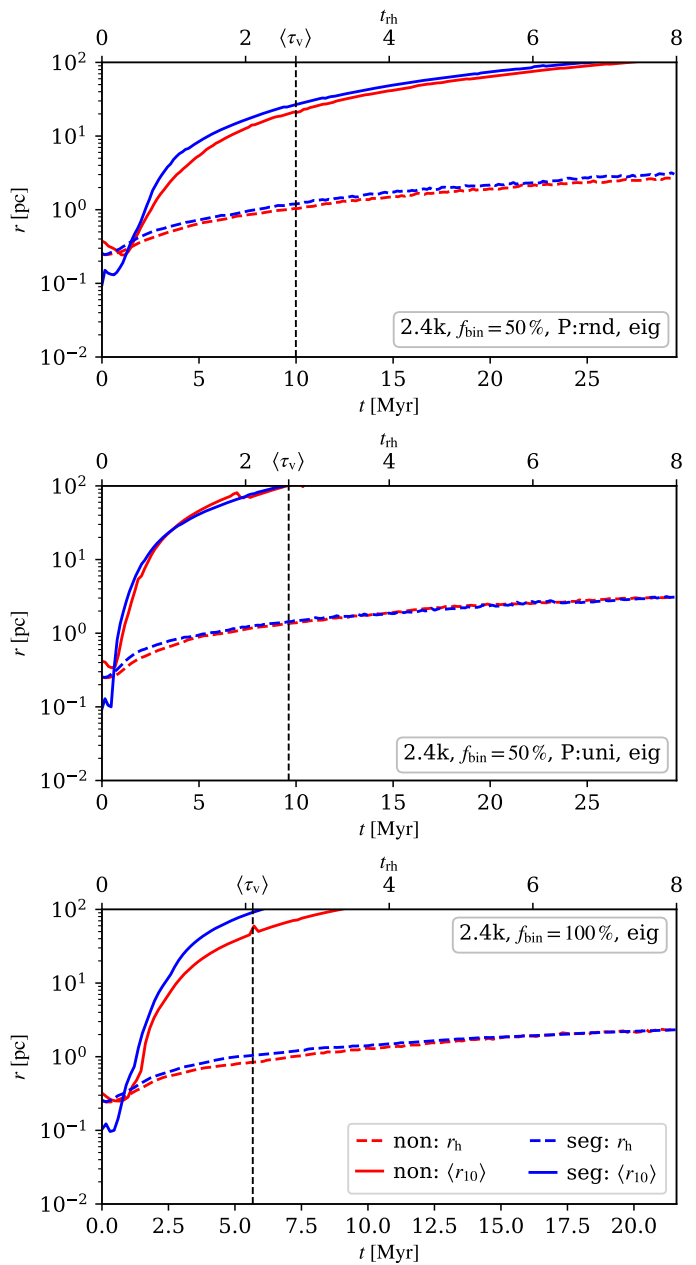

        \centering
        \includegraphics[width=\linewidth]{{{bin_seg_0-top_10_mass_avg-eig}}}

        \includegraphics[width=\linewidth]{{{bin_seg_3-top_10_mass_avg-eig}}}

        \includegraphics[width=\linewidth]{{{bin_100-top_10_mass_avg-eig}}}
        \caption{Average positions of the 10\.\% most massive stars in each model. For reference, the half-mass radius is plotted with a dashed colour line, and the mean time $\langle\tauv\rangle$ of each model is plotted by a dashed vertical line.} 
        \label{fig:top_mass}
\end{figure}

Stellar evolution relates to mass loss from stars, hence weakening of the potential well of the cluster and subsequent expansion of the system. Moreover, heavy stars evolving towards black holes (BHs) or neutron stars (NSs) could go through a supernova, receiving a non-negligible natal kick \citep[e.g.][]{kicks_ns,bh_ns_kicks} and leave the cluster \citep[cf.][]{baumgardt_sollima,pavl_bh}. Based on the IMF \citep{kroupa} and the stellar evolutionary algorithm by \cite{sse}, only five stars had the mass required to become a BH, and another $\approx 30$ could become a NS. However, at the time $\langle\tauv\rangle,$ which is close to 12\,Myr, only the BHs would have been fully evolved, assuming sub-solar metallicity \citep[according to the estimate of][]{pavl_bh}; using the solar metallicity, the evolution would take even longer. Thus, by the time $\langle\tauv\rangle$, the cluster could lose at most $0.1\,\mcl$ if all of the BHs escaped. In our models, however, the most massive stars (and all five potential BHs) are put into binaries from the beginning. Those binaries survived for several Myr, which means that they would also evolve a significant part of their life according to binary stellar evolution algorithms \citep[e.g.][]{bse} including also the phase of common envelope -- redistribution of mass may rejuvenate them and delay the supernova explosion to later times than, for example, the estimated 12\,Myr \citep[cf.][]{1992ApJ...391..246P,2003NewA....8..817D,zapartas2017,pavl_bh,dicarlo}. In addition to that, we must also take into account the inner dynamical evolution. As it is visible in Fig.~\ref{fig:top_mass}, some of the top 10\,\% most massive stars escape due to three- or multi-body encounters regardless of their stellar evolution. A similar result in the sense of binary BH ejection was also reported by \citet{dicarlo}. It appears, therefore, that in our models, including stellar evolution should not change the initial evolution or the time $\langle\tauv\rangle$. Nevertheless, simulations with full treatment for single- and binary-stellar evolution (possible with different evolutionary algorithms) would be necessary to fully determine this.

\begin{figure*}[!htb]
\begin{minipage}[t]{.49\linewidth}
        \centering
        \includegraphics[width=\linewidth]{{{bin_seg_0-rl-eig}}}

        \includegraphics[width=\linewidth]{{{bin_seg_3-rl-eig}}}

        \includegraphics[width=\linewidth]{{{bin_100-rl-eig}}}
        \caption{Evolution of Lagrangian radii in presented models with eigenevolution. The Lagrangian radii are defined as the radii of concentric spheres containing fixed mass fractions (printed to the right of each curve), where the core is identified as the cluster's density centre \citep[cf.][]{casertano_hut,aarseth}. The mean value $\langle\tauv\rangle$ of each model is plotted by a dashed vertical line for reference.}
        \label{fig:lagr}
% \end{figure}

\end{minipage}
\hfill %\hspace{2pc}
\begin{minipage}[t]{.49\linewidth}

% \begin{figure}
        \centering
        \includegraphics[width=\linewidth]{{{bin_seg_0-rl-tide-eig}}}

        \includegraphics[width=\linewidth]{{{bin_seg_3-rl-tide-eig}}}

        \includegraphics[width=\linewidth]{{{bin_100-rl-tide-eig}}}
        \caption{Evolution of the half-mass radius (dashed colour lines) of each model calculated from stars that are bound to the cluster, i.e.\ up to the tidal radius from Eq. \eqref{eq:tidal}, which is plotted by a solid line. An artificial half-mass radius, which should represent the cluster evolution after gas expulsion, is plotted by the lighter dashed line for each model. The mean value $\langle\tauv\rangle$ of each model is plotted by a dashed vertical line for reference.}
        \label{fig:lagr_tide}
        
\end{minipage}

\end{figure*}

Modelling an isolated star cluster is also an idealisation. Due to the tidal field, a star cluster would gradually lose its stars through evaporation and shrink.
An estimate of this tidal radius is
\begin{equation}
  \label{eq:tidal}
  r_\mathrm{t} = r_\mathrm{G} \left( \frac{M_\mathrm{ct}}{3 M_\mathrm{G}} \right)^{\frac{1}{3}} \,,
\end{equation}
where $M_\mathrm{ct}$ is the mass of the cluster within this radius, $r_\mathrm{G}$ is the distance of the cluster from the centre of the Galaxy, and $M_\mathrm{G}$ is the mass of the Galaxy comprised in the radius $r_\mathrm{G}$ \citep[e.g.][]{binney_tremaine}.

In Fig.~\ref{fig:lagr}, we plot the evolution of our isolated star clusters by the means of the Lagrangian radii. As expected, the cluster goes through the core collapse (at $\tcc \approx 0.25\,\trel$; estimated from the minimum of the innermost radii) and then expands -- a measure of the cluster expansion could be the half-mass radius. In Fig.~\ref{fig:lagr_tide}, we plot the evolution of the half-mass radius calculated from the stellar mass contained up to the cut-off at the tidal radius -- we assume $r_\mathrm{G} = 5\,\kpc$, which yields $M_\mathrm{G} \approx 5\times10^{10}\,\msun$ \citep{galaxy_mass,gallaxy_mass_new}.
As the tidal radius and the upper boundary for evaluating mass segregation are the same throughout the integration ($r_\mathrm{t} \approx r_{\nbin} = 10\,\pc$), it is unlikely that the resulting time $\langle\tauv\rangle$ would change if the clusters were placed in the background potential of the Galaxy.

The most important simplification that we used is that the clusters do not contain gas. Gas expulsion due to radiation pressure from massive stars is responsible for an early cluster expansion. In the case of the \object{ONC}, the cluster could inflate on average by a factor of $r_{\rm scl} \approx 3$ between approximately 0.5 and 1\,Myr \citep[e.g.][]{subr}. In Fig.~\ref{fig:lagr_tide}, we also plot an artificial half-mass radius represented by lighter colours. Up to 0.75\,Myr, it is equal to the half-mass radius calculated from the cluster mass within the tidal radius, and from 0.75\,Myr onward, it is scaled up three times to simulate gas expulsion. Since the separations of stars in the cluster become also about three times larger after gas expulsion, the evolution of the whole system slows down by a factor of $r_{\rm scl}^{1.5} \approx 5.2$, which is also included in the modified half-mass radius in Fig.~\ref{fig:lagr_tide}. All half-mass radii in Fig.~\ref{fig:lagr_tide} (modified or not) reach a similar value at 5\,Myr and then evolve almost identically. Nonetheless, we conjure that the time $\langle\tauv\rangle$ would shift to later times in models with gas expulsion due to their slower evolution. Models of embedded clusters with gas should, however, be calculated to clarify this.

\end{document}